\documentstyle[12pt]{article}
\advance\textheight by \topskip
\begin{document}
\vspace{-0.2cm}
\begin{flushright}
CINVESTAV--GRG--94/12\\  May 94\\ (Revised: March 95)\\
\vskip2mm
hep-th/9504155
\end{flushright}
\vskip1cm
\begin{center}
{\LARGE \bf Symmetries of the stationary\\ \vskip 3mm
Einstein--Maxwell--dilaton Theory}
\end{center}
\begin{center}
{\bf D.V. Gal'tsov,\footnote{On leave from
the Dept.\ of Theoretical Physics, Moscow State
University, 119899 Moscow, Russia; e--mail: galtsov@grg.phys.msu.su}
\footnote{Supported by CONACyT--Mexico. } A.A. Garc\'\i a,
and O.V. Kechkin$^{1\;2}$}\\
\vskip1cm
{\em Centro de Investigaci\'on y de Estudios Avanzados del I.P.N.\\
Departamento de F\'\i sica,
Apdo. Postal 14-740, 07000,
M\'exico, D.F., M\'exico}
\end{center}
\vskip 1 cm

\centerline{\bf Abstract}

\begin{quotation}
\vskip .5 cm
Gravity coupled three--dimensional $\sigma$--model describing the stationary
Einstein--Maxwell--dilaton system with general dilaton coupling is studied.
Killing equations for the corresponding five--dimensional target space are
integrated. It is shown that for general
coupling constant $\alpha$ the symmetry algebra is
isomorphic to the maximal solvable subalgebra of
$sl(3,R)$.  For two critical values
$\alpha =0$ and $\alpha =\sqrt{3}$, Killing algebra
enlarges to the full $sl(3,R)$ and $su(2,1)\times R$ algebras
respectively, which correspond
to five--dimensional Kaluza--Klein and four--dimensional
Brans--Dicke--Maxwell theories. These two
models are analyzed in terms of the unique real variables.
Relation to the description in terms of complex
Ernst potentials  is discussed.
Non--trivial discrete maps between  different subspaces of the
target space are found and used to generate new arbitrary--$\alpha$ solutions
to dilaton gravity.

\end{quotation}

\newpage

\section{Introduction}
\renewcommand{\theequation}{1.\arabic{equation}}
\setcounter{equation}{0}

Hidden symmetries in gravity and supergravity theories dimensionally reduced
to three and two dimensions were extensively studied earlier \cite{nk},
\cite{ki}, \cite{ms}, \cite{bgm}, \cite{ju}, \cite{ex}. As it is well--known,
vacuum Einstein equations in a space--time of arbitrary dimensions possessing
a sufficient number of Killing vectors to make the system effectively
three--dimensional, can be presented in the form of three--dimensional
gravity coupled $\sigma$--model with a symmetric target space. When further
reduced to two dimensions via an imposition of an additional Killing
vector field, the system becomes integrable by means of the inverse
scattering transform method \cite{bz} \cite{he} \cite{br}. Similar property
is shared by the four--dimensional Einstein--Maxwell
theory \cite{ki}, \cite{al},
\cite{ma}, \cite{egk}, as well as by some more general gravity--coupled
scalar--vector models resulting from supergravities \cite{bgm}, \cite{ju}.

More recent interest to this subject is related to the search of lower
dimensional string models. It turns out to be useful to study exact
solutions of the low--energy effective string actions as the first
step in the  search of exact string backgrounds.
 Surprisingly, many of exact solutions were obtained in
the closed analytical form
\cite{gi}, \cite{gm}, \cite{ghs}. Moreover, certain similarity with
corresponding solutions to the vacuum Einstein equations can be observed.
This fact indicates on  the existence of hidden symmetries of
lower--dimensional reductions of the theory. It was shown recently that
within the context of the Einstein--Maxwell--dilaton--axion theory the
hidden symmetry group includes Ehlers--Harrison--type transformations
\cite{gk}, the corresponding two--dimensional truncation leads
to integrable theory \cite{g}. It was observed also that purely
dilatonic gravity (without an axion) is less symmetric apart form
two exceptional values of the dilaton coupling constant.
It is the purpose of the present paper to discuss this subject in some
details.

We consider an Einstein--Maxwell--dilaton (EMD) system with
general dilaton coupling in four space--time dimensions
\begin{equation}
S=\frac{1}{16\pi}\int \left(-R+2(\partial\phi )^2-
e^{-2\alpha\phi}F^2\right)\sqrt{-g}\;d^4x,
\end{equation}
where $\phi$ is the real scalar field (dilaton), $F=dA$ is the
Maxwell two--form, $\alpha$ is the dilaton coupling constant.
This theory was suggested as one of stringy
gravity models \cite{ghs}. It is also interesting
as a minimal model continuously  interpolating betweeen
two highly symmetric systems: Einstein--Maxwell (EM) and five--dimensional
Kaluza--Klein (KK) theories.

For $\alpha =0$ the action (1.1) describes the EM system
with the gravitationally coupled scalar field.
It is well--known that the pure EM system
becomes an integrable theory provided a two--dimensional Abelian Killing
space--time symmetry is imposed.
Possible way to demonstrate
this property consists in the following. Imposing first one Killing vector
field one can dimensionally reduce the system to a gravity coupled
three--dimensional sigma--model with
the target space   $SU(2,1)/S(U(1)\times U(2))$
\cite{ki}, \cite{ma}, \cite{egk}.
Such a system in presence of the second Killing vector field
commuting with the first one  admits  two--dimensional modified
chiral matrix representation of Belinskii--Zakharov type \cite{bz}.
With a scalar field added, the $\alpha =0$ action (1.1) can
equivalently be thought of as the Brans--Dicke--Maxwell (BDM) model in the
Einstein frame (with the Brans--Dicke parameter $\omega =-1$). Obviously
this system posesses the same integrability properties as the EM one
\cite{gam}.

Similar integrability property is shared by the KK theory with
three commuting Killing vectors, one
of which corresponds to the $x^5$--translations, \cite{ne}, \cite{ms},
\cite{mt}.
In the adapted system of coordinates this
theory can be presented as the four--dimensional EMD
system (1.1) with the coupling constant $\alpha=\sqrt{3}$.
With two commuting Killing symmetries imposed,  it
reduces to the $SL(3,R)/SO(3)$ modified chiral matrix model.

Both $SU(2,1)$ and $SL(3,R)$ groups are eight--parametric,
so it is natural to investigate a possibility
of a deeper link between the stationary EM and KK theories.
Clearly, the action (1.1)  is the simplest model which ensures
a continuous interpolation. Using the
real parametrization of the target space corresponding to the stationary
reduction of the action (1.1) one can make explicit the relationship
between these two structures.
{}From the point of view of the string theory, the distinguished value of
the dilaton coupling is $\alpha =1$. So apart from
the purely mathematical question about the correspondence between KK and
BDM chiral models, it is important to know whether the stringy MD
model shares the same integrability property.

The rest of the paper is organized as follows. In Sec. 2
three--dimensional
sigma--model representation is derived
for the stationary EMD system with an arbitrary dilaton coupling constant.
In Sec. 3 we present a detailed investigation of isometries of the
corresponding target space. The nature of Killing algebra for
non--critical
coupling is discussed in Sec. 4.  Two particular cases $\alpha =\sqrt{3}$ and
$\alpha =0$ are then considered in detailes
using the real target space variables (Sec. 5, 6).  For the BDM case
a correspondence between the symmetry generators
obtained and those known in
conventional terms of the Ernst potentials is established (Sec. 6). In
Sec. 7 we describe complex descrete transformations of the EMD system similar
to Bonnor transformations in the EM theory. They are used
to derive new asymptotically flat solutions to dilaton gravity describing
dipole field configurations.
We conclude with some remarks concerning the nature of the symmetry
breaking by a non--critical dilaton.

\section{  Dimensional Reduction.}
\renewcommand{\theequation}{2.\arabic{equation}}
\setcounter{equation}{0}

 Assuming four--dimensional metric in (1.1) to admit a time--like Killing
vector field, one can write the space--time line element as
\begin{equation}
ds^2=f(dt-\omega_idx^i)^2-f^{-1}h_{ij}dx^idx^j,
\end{equation}
where the function $f$, the one--form $\omega =\omega_idx^i$ and the 3--metric
$h_{ij}$
depend only on the space coordinates $x^i, i=1, 2, 3,$.
It can be easily shown
that the corresponding Maxwell field is fully describable in terms of two
real--valued
functions $v$ and $a$ of $x^i$ exactly as in the case of the pure EM field
\cite{iw}.
Indeed, Maxwell equations and  Bianchi identities following from (1.1) read
\begin{equation}
\partial_\nu(\sqrt{-g}e^{-2\alpha\phi}F^{\mu\nu})=0,
\end{equation}
\begin{equation}
\partial_\nu(\sqrt{-g}{\tilde F}^{\mu\nu})=0,
\end{equation}
where ${\tilde F}^{\mu\nu}=\frac{1}{2}E^{\mu\nu\lambda\tau}F_{\lambda\tau}, \;
E^{\mu\nu\lambda\tau}=\epsilon^{\mu\nu\lambda\tau}/\sqrt{-g}.$ With the
assumption of stationarity, the $\mu =i$ component of (2.3) is satisfied
by the substitution
\begin{equation}
F_{i0}=\frac{1}{\sqrt{2}}\partial_iv,
\end{equation}
while the $\mu =i$ component of the Eq. (2.2) is solved by
\begin{equation}
F^{ij}=\frac{f}{\sqrt{2h}}e^{2\alpha\phi}\epsilon^{ijk}\partial_ka.
\end{equation}
\noindent
The quantities $v$ and $a$ may be regarded as electric and magnetic potentials
respectively.  The remaining components of the $F^{\mu\nu}$ can be
expressed in terms of $v$ and $a$ using the relation \cite{iw}
\begin{equation}
F^{i0}=F^{ij}\omega_j-h^{ij}F_{j0},
\end{equation}
where $h^{ij}$ is the 3-inverse of $h_{ij}.$ Another useful relation is
\begin{equation}
F_{ij}=f^{-2}h_{ik}h_{jl}F^{kl}+2F_{0[i}\omega_{j]}.
\end{equation}

Following Israel and Wilson \cite{iw} one can introduce a 3--dual
to the rotation 2--form $d\omega$
\begin{equation}
\tau^i=-f^2\frac{\epsilon^{ijk}}{\sqrt{h}}\partial_j\omega_k,
\end{equation}
which is invariant under the time transformation $t\rightarrow t+T(x^i).$  We
assume further that the indices of all 3--dimensional quantities are
raised and lowered with the 3--metric $h_{ij}$ while for 4--dimensional
tensors one still uses $g_{\mu\nu}$.
Then the relevant components of
the 4--Ricci tensor$R_{\mu\nu}$ (defined as
$R_{\mu\nu}=\partial_\mu\Gamma^\alpha
_{\nu\alpha}-...)$ can be presented as
\begin{equation}
R_{00}=\frac{1}{2}\left(f\Delta f-(\nabla f)^2+\tau^2 \right),
\end{equation}
\begin{equation}
R^i_0=\frac{f}{2\sqrt{h}}\epsilon^{ijk}\tau_{k,j},
\end{equation}
\begin{equation}
R^{ij}=f^2\Re^{ij}-\frac{1}{2}\left[(\nabla^if)(\nabla^jf)+\tau^i\tau^j
-h^{ij}\left(f\Delta f-(\nabla f)^2+\tau^2\right)\right],
\end{equation}
where $\Re^{ij}$ is the Ricci tensor of the 3--space, $\nabla_i$ denotes
3--covariant derivative, $\Delta =\nabla^2$, and 3--vector scalar products are
understood with respect to the metric $h_{ij}.$

The corresponding components of the stress--energy tensor are:
\begin{equation}
16\pi (T_{00}-\frac{1}{2}g_{00}T)=f\left((\nabla v )^2e^{-2\alpha\phi }+
(\nabla a)^2e^{2\alpha\phi}\right),
\end{equation}
\begin{equation}
8\pi T_0^i=\frac{f}{\sqrt{h}}\epsilon^{ijk}(\nabla_jv)(\nabla_ka),
\end{equation}
\[
8\pi(T^{ij}-\frac{1}{2}g^{ij}T)=-f\left(e^{-2\alpha\phi}(\nabla^iv)(\nabla^jv)+
e^{2\alpha\phi}(\nabla^ia)(\nabla^jv)\right)
\]
\begin{equation}
+\frac{fh^{ij}}{2}\left(e^{-2\alpha\phi}(\nabla v)^2+e^{2\alpha\phi}(\nabla
a)^2\right)+
2f^2(\nabla^i\phi)(\nabla^j\phi )
\end{equation}
Note that the dilation does not influence mixed components of the Einstein
equations.  Hence, comparing (2.10) and (2.13) one obtains
\begin{equation}
\tau_i=w_i+\nabla_i\chi,
\end{equation}
where
\begin{equation}
w_i=v\nabla_ia-a\nabla_iv,
\end{equation}
and $\chi$ is the twist potential defined up to an additive constant.  In
terms of $\chi$ and {\bf w} the 00--component of the Einstein equations
will read
\begin{equation}
f\Delta f-(\nabla f)^2=f\left((\nabla v)^2e^{-2\alpha\phi}+(\nabla a)^2
e^{2\alpha\phi}\right)-(\nabla\chi + {\bf w})^2.
\end{equation}
The divergence of (2.15) combined with (2.8) leads to the equation for
$\chi$
\begin{equation}
f\Delta\chi -2\nabla f(\nabla\chi +{\bf w})+f(v\Delta a-a\Delta v)=0.
\end{equation}
To obtain second order equations for $v$ and $a$ one has to use the $\mu =0$
components of the equations (2.2), (2.3).  Then taking into account
(2.6), (2.7), (2.8) and (2.15) one gets
\begin{equation}
f^2\nabla (f^{-1}e^{-2\alpha\phi}\nabla v)+(\nabla\chi +{\bf w})\nabla a=0,
\end{equation}
\begin{equation}
f^2\nabla (f^{-1}e^{2\alpha\phi}\nabla a)-(\nabla\chi +{\bf w})\nabla v=0.
\end{equation}
Finally, the dilaton equation in terms of the same variables reads
\begin{equation}
2f\Delta\phi =\alpha \left(e^{-2\alpha\phi}(\nabla v)^2-e^{2\alpha\phi}(\nabla
a)^2\right).
\end{equation}

The set of equations (2.17)--(2.21) has to be completed by the remaining
$ij$--Einstein equations.  Combining (2.11) and (2.14) and using (2.15)
one obtains for the 3--dimensional Ricci tensor the following expression
\[
\Re_{ij}=\frac{1}{2f^2}\left[(\nabla_i f)(\nabla_jf)+(\nabla_i\chi + w_i)
(\nabla_j\chi +w_j)\right]+
\]
\begin{equation}
+2(\nabla_i\phi )(\nabla_j\phi )-f^{-1}\left[e^{-2\alpha\phi}(\nabla_i v)
(\nabla_jv)+e^{2\alpha\phi}(\nabla_ia)(\nabla_j a)\right].
\end{equation}
The system (2.17)--(2.22) provides a fully 3--dimensional description of the
stationary EMD system with an arbitrary dilaton coupling constant $\alpha$.
It can be regarded as the 3--dimensional Einstein--matter  system with five
real scalar fields
\begin{equation}
\varphi^A=(f,\chi,v,a,\phi ),\quad A=1,...,5,
\end{equation}
acting as a source.  It can equivalently be derived from the following
3--dimensional gravity coupled $\sigma$--model action
\begin{equation}
S_{\sigma}=\int \left(\Re -{\cal G}_{AB}(\varphi )
\partial_i\varphi^A\partial_j\varphi^B\right)h^{ij}\sqrt{h}\;d^3x,
\end{equation}
where $\Re = \Re^i_i,$
and ${\cal G}_{AB}(\varphi )$ is the target space metric
\[
{\cal G}={\cal G}_{AB}d\varphi^Ad\varphi^B=\frac{1}{2f^2}
\left(df^2+(d\chi+vda-adv)^2\right)-
\]
\begin{equation}
-\frac{1}{f}(e^{-2\alpha\phi}dv^2+e^{2\alpha\phi}da^2)+2d\phi^2.
\end{equation}
For $\alpha =0$ and $\phi = const$ this metric reduces to one given by
Neugebauer and Kramer for the EM system \cite{nk}. It is worth to be noted
that the dilaton equation (2.21) for  $\phi = const$ gives a constraint
on the Maxwell field, $F^2=0$. Clearly the stationary EMD system
may only have $\phi = const$ solutions
if $F^2=0$. Hence, the EM system {\em is not}
a particular case of the EMD system. Rather, when we put $\alpha =0$,
we get the BMD system, this case will be discussed in details in the Sec. 6.

This  representation of the stationary EMD system
will be the starting point for the subsequent investigation of the associated
hidden symmetries.  Some of them can be readily found from
the explicit expression
(2.25) for the target space metric.  However, it turns out that the number
of symmetry generators depends on the value of the dilaton coupling constant
$\alpha$ in a somewhat  tricky way.  One needs to undertake the complete
analysis of the target space isometries in order to understand mutual
relationship between different symmetry groups arising for
two critical values of $\alpha$.
\section{ Integration of the target space Killing equations}
\renewcommand{\theequation}{3.\arabic{equation}}
\setcounter{equation}{0}

It is convenient to introduce instead of $f$ and $\phi$
the following new variables
\begin{equation}
\eta =\alpha\phi -\frac{1}{2}ln f, \quad
\xi =-(\alpha\phi +\frac{1}{2}ln f),
\end{equation}
and to define parameters
\begin{equation}
p=\frac{\alpha^2+1}{2\alpha^2}, \quad  q=\frac{\alpha^2-1}{2\alpha^2},
\end{equation}
provided $\alpha\not= 0$ (the case $\alpha =0$ will be discussed separately).
Then the target space metric (2.25) takes the simple form
\begin{equation}
{\cal G} =p(d\eta^2+d\xi^2)+2qd\eta d\xi -e^{2\xi} dv^2-e^{2\eta}da^2
+\frac{1}{2}e^{2(\eta +\xi )}(d\chi +vda-adv)^2.
\end{equation}

Our aim is to find all isometries of the target space, that is to construct a
complete set of solutions to the Killing equations
\begin{equation}
X_{A;B}+X_{B;A} =0,
\end{equation}
where covariant derivatives refer to the metric $\cal G$.
Contracting (3.4) with $d\varphi^Ad\varphi^B$ and substituting (3.3) one
obtains
the following equation in terms of bilinear forms:
\[
\frac{1}{2}(d\chi +vda-adv)[(X^\eta +X^\xi )(d\chi +vda-adv)+
X^vda-X^adv+dX^\chi
\]
\[
+vdX^a-adX^v ] e^{2(\eta+\xi )}+p(dX^\eta d\eta +dX^\xi d\xi )+q
(dX^\xi d\eta +dX^\eta d\xi )-
\]
\begin{equation}
-e^{2\eta} (X^\eta da+dX^a)da-e^{2\xi}(X^\xi dv+dX^v)dv=0.
\end{equation}

It contains a set of 15 independent equations, which can be solved
by extracting the explicit dependence of $X^A$ on the field
variables $(\eta, \xi, \chi, v, a)$
step by step.
Collecting the $d\eta^2$ and $d\xi^2$ terms in (3.5), one obtains two
equations
\begin{equation}
(pX^\eta +qX^\xi )_{,\eta}=0 \;\; , \;\; (pX^\xi + qX^\eta )_{,\xi}
=0,
\end{equation}
which can be solved
\begin{equation}
X^\eta =\frac{p\pi -q\kappa}{p-q} \;\;\; , \;\;\; X^\xi =
\frac{p\kappa - q\pi}{p-q},
\end{equation}
in terms of two functions of 4 variables $\pi =\pi (\xi, \chi, v, a)$
and $\kappa =\kappa (\eta, \chi, v, a).$  Then
from the $d\xi d\eta$ equation
\[
p({X^\eta}_{,\xi} +{X^\xi}_{,\eta} ) + q({X^\xi}_{,\xi} + {X^\eta}_{,\eta} )=0,
\]
\noindent
one can find an explicit dependence of $\pi$ and $\kappa$ on $\xi$
and $\eta$
\[
\pi =A(\chi, v, a)\xi +B(\chi, v, a),
\]
\begin{equation}
\kappa =-A(\chi , v, a)\eta +C(\chi , v, a).
\end{equation}
(Here and in what follows capital Latin letters denote the differentiable
functions of variables indicated in the parenthesis.)

To extract the $\chi$--dependence we use the $d\eta d\chi$ and
$d\xi d\chi$ equations
\begin{equation}
2(pX^\eta+qX^\xi )_{,\chi} +e^{2(\eta +\xi )}(X^\chi +vX^a-aX^v)_{,\eta}=0,
\end{equation}
\begin{equation}
2(pX^\xi +qX^\eta )_{,\chi}+e^{2(\eta +\xi )}(X^\chi +vX^a-aX^v)_{,\xi}=0,
\end{equation}
yielding
\[
A=A(v,a),\quad C=B+F(v,a),
\]
\begin{equation}
X^\chi+vX^a-aX^v=e^{-2(\eta+\xi )}B_{,\chi}+D(\chi,v,a),
\end{equation}
together with the $d\chi^2$ equation
\begin{equation}
X^\eta + X^\xi +(X^\chi +vX^a -aX^v)_{,\chi}=0.
\end{equation}
They imply
\[
A\equiv 0, \;\;\; B=G(v, a) \chi +H(v, a),
\]
\begin{equation}
D=-G(v, a)\chi^2-\left(2H(v, a)+F(v, a)\right)\chi +I(v, a).
\end{equation}

The next step of the derivation is the $a, v$--reduction.
{}From the $dad\eta$ term in (3.5)
one gets
\begin{equation}
p\chi^\eta_{,a}+\frac{v}{2}e^{2(\eta +\xi
)}(X^\chi+vX^a-aX^v)_{,\eta}-e^{2\eta}
X^a_{,\eta}+qX^\xi_{,a}=0.
\end{equation}
{}From here, with account for (3.10) and (3.13), one arrives at
\begin{equation}
X^a=-\frac{1}{2}e^{-2\eta}(G_{,a}\chi +H_{,a}-Gv)+K(\xi,\chi, v, a).
\end{equation}
Then, from the $dad\xi$ equation
\begin{equation}
(pX^\xi +qX^\eta)_{,a}+\frac{v}{2}e^{2(\eta +\xi
)}(X^\chi+vX^a-aX^v)_{,\xi}-e^{2\eta}X^a_{,\xi}=0
\end{equation}
the function K is seen to be independent on $\xi,\; G = G(v)$ and also
\begin{equation}
H+F =G(v)av +L(v).
\end{equation}
Similarly, using the $dvd\xi$ equation
\begin{equation}
pX^\xi_{,v}-\frac{a}{2}e^{2(\eta +\xi )}(X^\chi+vX^a-aX^v)_{,\xi}-e^{2\xi}
X^v_{,\xi}+qX^\eta_{,v}=0,
\end{equation}
we find
\begin{equation}
X^v=-\frac{1}{2}e^{-2\xi}(G_{,v}\chi+H_{,v}+F_{,v}+Ga)+M(\eta ,\chi , v, a).
\end{equation}
Entering with (3.19) into the $dvd\eta$ equation one gets
\begin{equation}
pX^\eta_{,v}-\frac{a}{2}e^{2(\eta +\xi )}(X^\chi+vX^a-aX^v)_{,\eta}
-e^{2\xi}X^v_{,\eta}+qX^\xi_{,v}=0.
\end{equation}
{}From here it can be derived
\[
G=const, \;\;\; H=-Gva+N(a),\;\;\;\; M=M(\chi, v, a),
\]
\begin{equation}
F=2Gva + L(v)-N(a).
\end{equation}

Collecting all previous results one can write the general solution of the
Killing equations as follows
\[
X^\eta =G\chi+\left(pN(a)-qL(v)-Gva\right)(p-q)^{-1},
\]
\[
X^\xi =G\chi+\left(pL(v)-qN(a)+Gva\right)(p-q)^{-1},
\]
\begin{equation}
X^v=-\frac{1}{2}e^{-2\xi}\left(2Ga+L'(v)\right)+M(\chi, v, a),
\end{equation}
\[
X^a=\frac{1}{2}e^{-2\eta}\left(2Gv-N'(a)\right)+K(\chi, v, a),
\]
\[
X^\chi +vX^a-aX^v=G(e^{-2(\eta +\xi )}-\chi^2)-\left(N(a)+L(v)\right)\chi
+I(v,a).
\]
with $G=const$, where primes stand for the derivatives with respect
to the corresponding single argument.

For a subsequent reduction one uses first the $da d\chi$ equation obtaining the
relations
\[
K=-\frac{1}{2}\left(L'(v)+2Ga\right)\chi +P(v, a),
\]
\begin{equation}
M=\frac{1}{2}\left(N'(a)-2Gv\right)\chi -\frac{1}{2}[I_{,a}+v(N+L)],
\end{equation}
and then the $dvd\chi$ equation giving
\[
M=\frac{1}{2}\left(N'(a)-2Gv\right)\chi + Q(v, a),
\]
\begin{equation}
K=-\frac{1}{2}\left(L'(v)+2Ga\right)\chi
+\frac{1}{2}\left(I_{,v}-a(N+L)\right).
\end{equation}
Combining (3.23) and (3.24) one gets
\[
P(v, a)=\frac{1}{2}[I_{,v}-a(N+L)],
\]
\begin{equation}
G(v, a)=-\frac{1}{2}[I_{,a}+v(N+L)].
\end{equation}
Furthermore, the $da^2$ equation yields $N=Ra+S$ with constant $R$ and $S$,
and
\begin{equation}
I_{,va}=(p-q)^{-1}(L-N+4qGav)+aN'-vL'.
\end{equation}
Similarly, from the $dv^2$ equation one gets $L=Tv+U$, with constant $T$ and
$U$, as well as (3.26) again.
Finally, the crossed term $dvda$ in
(3.5) gives two additional equations for $I(v, a)$
\begin{equation}
I_{,vv}=2a(Ga+T),
\end{equation}
\begin{equation}
I_{,aa}=2v(Gv-R).
\end{equation}
\indent Differentiating (3.26) over $v$ and (3.27) over $a$ we obtain the
following
consistency condition:
\begin{equation}
\frac{q}{p-q}(T+2Ga)=T+2Ga.
\end{equation}
In the same way, differentiating (3.26) over $a$ and (3.28) over $v$ one
gets
\begin{equation}
\frac{q}{p-q}(2Gv-R)=2Gv-R.
\end{equation}
Now it is clear from (3.29) and (3.30), that the constants $G, R, T$ can
be non-zero if and only if
\begin{equation}
\frac{q}{p-q}=1,
\end{equation}
what, on account for (3.2), means
\begin{equation}
\alpha=\sqrt{3}.
\end{equation}
For all other (non--zero) values of $\alpha$ one has $G=R=T=0,$ and the
solution of (3.26)--(3.28) will read
\begin{equation}
I=\frac{1}{p-q}(U-S)av+Wa+Vv+Z,
\end{equation}
where $W, V$ and $Z$ are constants, and $N=S, L=U.$
Therefore for an arbitrary value of $\alpha$, except for $\alpha =0$ and
$\alpha =\sqrt{3}$, the general solution of the Killing equations obtained
by substituting
(3.24), (3.25), (3.33) into (3.22) reads
\[
X=X^A\frac{\partial}{\partial\varphi^A}=S(p-q)^{-1}
\left[p(\partial_\eta-a\partial_a) +
q(v\partial_v-\partial_\xi)\right] -S\chi\partial_\chi+
\]
\[
+U(p-q)^{-1}\left[p(\partial_\xi -v\partial_v)+q
(a\partial_a-\partial_\eta)-\chi\partial_\chi\right]
+\frac{1}{2}V(\partial_a+v\partial_\chi)+
\]
\begin{equation}
+\frac{1}{2}W(a\partial_\chi-\partial_v)+Z\partial_\chi.
\end{equation}
It describes five--parametric isometry group of the target space for any
non-critical $\alpha$.  For $\alpha =0$ the metric representation (3.3)
is not valid, this case will be discussed later in the Sec. 6.
For $\alpha =\sqrt{3}$ three additional Killing
vectors arise which correspond to the non--zero constants $G, R, T$.
\section{ Symmetry algebra for a non--critical coupling}
\renewcommand{\theequation}{4.\arabic{equation}}
\setcounter{equation}{0}

It is convenient to choose five independent Killing vectors as
follows:
\[
X_1=\partial_\eta -a\partial_a-\chi\partial_\chi ,
\]
\[
X_2=\partial_\xi -v\partial_v-\chi\partial_\chi,
\]
\begin{equation}
X_3=\partial_a+v\partial_\chi,
\end{equation}
\[
X_4=\partial_v-a\partial_\chi,
\]
\[
X_5=2\partial_\chi,
\]
where linear combinations of $S$ and $U$ used in (3.34) to get
$X_1$ and $X_2$ symmetric under the interchange $\eta\leftrightarrow\xi$.
$a\leftrightarrow v$.  Note that the discrete duality transformation
\begin{equation}
\eta\leftrightarrow\xi ,\;\;\; a\leftrightarrow v,\;\;\;
\chi\leftrightarrow -\chi,
\end{equation}
which  is obviously the symmetry of the target space metric (3.3), leaves the
whole set (4.1) invariant.
\indent

Five Killing vectors (4.1) form a closed 5--dimensional algebra X
\begin{equation}
[{X_\mu , X_\nu }]={C^\lambda}_{\mu\nu}X_\lambda ,\quad \mu ,\nu ,\lambda
=1,...,5,
\end{equation}
with the following non--zero structure constants:
\begin{equation}
{C^3}_{13}={C^4}_{24}={C^5}_{15}={C^5}_{25}=-{C^5}_{34}=1
\end{equation}
This algebra is solvable.  Indeed, its derivative $X'$ contains as basis
vectors $X_3, X_4,X_5,$  the second derivative is
one--dimensional $X^{''}=X_5$ , and we have the following chain of subalgebras
\begin{equation}
0=X^{'''}\subset X^{''}\subset X^{'}\subset X,
\end{equation}
where all terms are the subsequent ideals of the previous ones.

Using (4.4) it can be shown that the Killing--Cartan
bilinear form
\begin{equation}
{C^\alpha}_{\mu\beta}{C^\beta}_{\nu\alpha}\Omega^\mu\Omega^\nu =2(\Omega^1)^2+
2\Omega^1\Omega^2+(\Omega^2 )^2
\end{equation}
is non--degenerate only on 1--2 subspace.

Such algebras are known to admit a representation in terms of the upper
triangle matrices.  Consistently with (4.2) $X_1$ and $X_2$ can be choosen
diagonal.  Then the following 3$\times$3 representation
$X_\mu\rightarrow e_\mu$ holds
\[
e_1=\frac{1}{3} \left(\begin{array}{ccc}
2 & 0 & 0\\
0 & -1 & 0\\
0 & 0 & -1\\
\end{array} \right),\quad
e_2=\frac{1}{3} \left(\begin{array}{ccc}
1 & 0 & 0\\
0 & 1 & 0\\
0 & 0 & -2\\
\end{array} \right),\quad
e_3=\left(\begin{array}{ccc}
0 & 1 & 0\\
0 & 0 & 0\\
0 & 0 & 0\\
\end{array} \right),\quad
\]
\begin{equation}
e_4=\left(\begin{array}{ccc}
0 & 0 & 0\\
0 & 0 & -1\\
0 & 0 & 0\\
\end{array} \right),\quad
e_5=\left(\begin{array}{ccc}
0 & 0 & 1\\
0 & 0 & 0\\
0 & 0 & 0\\
\end{array} \right).
\end{equation}
Obviously, this set constitutes  a basis for the upper triangle subalgebra of
$sl(3,R).$

{}From (4.6) it is clear that, the algebra  of Killing vectors (4.3)
is insufficient to provide the
target space with a symmetric Riemannian space structure.
{}From 5 generators (4.1) 3 express pure gauge degrees of freedom ($X_3$ and
$X_4$---electromagnetic, $X_5$---gravitational).
The corresponding finite
transformations read respectively:
\begin{equation}
a\rightarrow a+\lambda_3 ,\;\;\; \chi\rightarrow \chi +\lambda_3 v,
\end{equation}
\begin{equation}
v\rightarrow v+\lambda_4 ,\;\;\; \chi\rightarrow \chi-\lambda_4 a,
\end{equation}
\begin{equation}
\chi\rightarrow\chi +\lambda_5 ,
\end{equation}
where $\lambda_3,\lambda_4,\lambda_5$ are the real group parameters. Two other
transformations $(X_1, X_2)$
are the linear combinations of a dilaton constant shift (accompanied by a
suitable
rescaling of $v$ and $a$) and a scale transformation:
\begin{equation}
\eta\rightarrow\eta +\lambda_1 \;\;\; a\rightarrow ae^{-\lambda_1}, \;\;\;
\chi\rightarrow\chi e^{-\lambda_1},
\end{equation}
\begin{equation}
\xi\rightarrow\xi +\lambda_2, \;\;\; v\rightarrow ve^{-\lambda_2}, \;\;\;
\chi\rightarrow\chi e^{-\lambda_2}.
\end{equation}

Comparing this with the EM case \cite{ki} we see that
the Ehlers--Harrison part of the symmetry group is
lacking.  This destroys the
integrability of the EM equations in presence of the second space--like
Killing vector field commuting with the initial time--like one is imposed.
Integrability property is,
however, restored for a particular value of the dilaton coupling constant
$\alpha =\sqrt{3}$, when the matrix algebra (4.7) enlarges to the full
$sl(3,R).$

\section{Kaluza--Klein theory}
\renewcommand{\theequation}{5.\arabic{equation}}
\setcounter{equation}{0}

 In this exceptional case the consistency conditions (3.29) and (3.30) are
fulfiled identically and the solution of the equations (3.26)--(3.28) reads
\begin{equation}
I=3(U-S)av+av(Tv-Ra+Gav)+Wa+Vv+Z.
\end{equation}
Thus, the general solution of the Killing equations contains 3 additional real
parameters: $G, R$ and $T$.  Substituting (4.1) into (3.23)--(3.25) and
further to (3.22), we get the 8--parametric solution.  In addition to  five
($\alpha$--independent) Killing vectors (3.35) now we have
\begin{equation}
X_6=v(\partial_\eta -2\partial_\xi )+(\frac{1}{2}e^{-2\xi}+ v^2)
\partial_v+\frac{1}{2}\chi_1\partial_a+\frac{1}{2}(ae^{-2\xi}+
v\chi_{-1})\partial_\chi ,
\end{equation}
\begin{equation}
X_7=a(\partial_\xi -2\partial_\eta )+(\frac{1}{2}e^{-2\eta}+a^2)\partial_a-
\frac{1}{2}\chi_{-1}\partial_v+\frac{1}{2}(a\chi_{1}-
ve^{-2\eta})\partial_\chi ,
\end{equation}
\[
X_8=-\frac{1}{2}(\chi_3\partial_\eta +\chi_{-3}\partial_\xi )+\frac{1}{2}
(v\chi_{-1}+ae^{-2\xi})\partial_v+\frac{1}{2}(a\chi_1-ve^{-2\eta})\partial_a+
\]
\begin{equation}
+\frac{1}{2}(a^2e^{-2\xi}+v^2e^{-2\eta}+a^2v^2+\chi^2-e^{-2(\xi +\eta )})
\partial_\chi,
\end{equation}
where $\chi_n\equiv\chi -nav.$  These generators were found by Neugebauer in
1969 \cite{ne} in slightly different variables
\footnote {The authors are grateful to Dr.~T.~Matos for
this reference.}.  Together with (4.1) they form the
$sl(3,R)$ algebra as it can be seen from
the commutation relations (4.3), now with
$\mu ,\nu ,\lambda =1,....,8.$  The set of non--zero structure constants
(4.4) is enlarged to include
\[
{C^7}_{17}={C^8}_{18}={C^6}_{26}={C^8}_{28}=-{C^6}_{38}={C^7}_{48}=-
{C^3}_{56}={C^4}_{57}=
\]
\[=-{C^8}_{67}=-{C^2}_{37}=-{C^1}_{46}={C^1}_{58}={C^2}_{58}=-1,
\]
\begin{equation}
{C^1}_{37}={C^2}_{46}=-2.
\end{equation}

New generators (5.2)--(5.4) have the following $3\times 3$ matrix counterparts
\begin{equation}
e_6= \left(\begin{array}{ccc}
0 & 0 & 0\\
0 & 0 & 0\\
0 & 1 & 0\\
\end{array} \right),\quad
e_7= \left(\begin{array}{ccc}
0  & 0 & 0\\
-1 & 0 & 0\\
0  & 0 & 0\\
\end{array} \right),\quad
e_8= \left(\begin{array}{ccc}
 0 & 0 & 0\\
 0 & 0 & 0\\
-1 & 0 & 0\\
\end{array} \right),
\end{equation}
completing the algebra of $sl(3,R)$.

The $sl(3,R)$ symmetry of the
2--stationary (admitting 2 commuting Killing vectors)
5--dimensional KK--theory
was first discovered by
Maison \cite{ms} using another formulation (see also \cite{mt}).  The
remarkable property of the target space (3.1) in the case $\alpha^2=3$ is
that it is a homogeneous symmetric Riemannian space on which the group
$SL(3,R)$ acts transitively.
It can be checked by a direct computation that the
Riemann tensor corresponding to the metric (3.1) for $p=2q=2/3$ is
covariantly constant: ${\nabla_AR^B}_{CDE}=0.$  Furthermore, the target space
is an Einstein space
\begin{equation}
R_{AB}=\frac{R}{5}{\cal G}_{AB},
\end{equation}
where the scalar curvature $R=-15.$  Maison \cite{ms} showed that this space
can be identified with
the coset $SL(3,R)/SO(3).$  Within the present framework this can be seen
as follows.

The Killing--Cartan metric constructed with the structure constants
(4.4) and (5.5),
\begin{equation}
\eta_{\mu\nu}=\frac{1}{12}{C^\alpha}_{\mu\beta}{C^\beta}_{\nu\alpha} =
\frac{1}{2}Tr(e_\mu e_\nu )
\end{equation}
has the following non--zero components:
\begin{equation}
\eta_{11}=\eta_{22}=2\eta_{12}=\frac{1}{3},\;\;\;
\eta_{37}=\eta_{46}=\eta_{58}=-
\frac{1}{2}.
\end{equation}
For the corresponding inverse tensor one finds
\begin{equation}
\eta^{11}=\eta^{22}=-2\eta^{12}=4,\;\;\; \eta^{37}=\eta^{46}=\eta^{58}=-2.
\end{equation}
Using these quantities, one can build the Killing one-forms
\begin{equation}
{\bf\tau}^\mu =\eta^{\mu\nu}X^A_\nu {\cal G}_{AB}d\varphi^B,
\end{equation}
which satisfy Maurer--Cartan equation
\begin{equation}
d{\bf\tau}^\mu
+\frac{1}{2}{C^\mu}_{\alpha\beta}{\bf\tau}^\alpha\wedge{\bf\tau}^\beta =0.
\end{equation}
Explicitly, ${\bf\tau}^\mu$ from (5.11) read
\[
{\bf\tau}^1=2d\eta +4ae^{2\eta}da-2ve^{2\xi}dv+\frac{1}{2}\chi_3{\bf\tau}^8,
\]
\[
{\bf\tau}^2=2d\xi+4ve^{2\xi}
dv-2ae^{2\eta}da+\frac{1}{2}\chi_{-3}{\bf\tau}^8,\]
\[
{\bf\tau}^3=2ad\eta +(2a^2e^{2\eta} +1 )da-\chi_{-1}e^{2\xi}dv+\frac{a}{2}
\chi_{-1}{\bf\tau}^8,
\]
\[
{\bf\tau}^4=2vd\xi
+(2v^2e^{2\xi}+1)dv+\chi_1e^{2\eta}da+\frac{v}{2}\chi_1{\bf\tau}^8,
\]
\[
{\bf\tau}^5=\chi_1d\eta +\chi_{-1}d\xi +(a+v\chi_{-1}e^{2\xi})dv-
\]
\[
-(v-a\chi_1e^{2\eta})da-\frac{1}{4}(a^2v^2-\chi^2+e^{-2(\xi +\eta
)}){\bf\tau}^8,
\]
\begin{equation}
{\bf\tau}^6=2e^{2\xi}dv-a{\bf\tau}^8,
\end{equation}
\[
{\bf\tau}^7=2e^{2\eta}da+v{\bf\tau}^8,\]
\[
{\bf\tau}^8=-2e^{2(\xi +\eta )}(d\chi +vda-adv).
\]

In terms of these Killing one--forms the line element of the target space (3.3)
with $p=2q=2/3$ can be written as
\[
{\cal G}=\frac{1}{2}\eta_{\mu\nu}{\bf\tau}^\mu\otimes{\bf\tau}^\nu=\frac{1}{6}
({\bf\tau}^1\otimes{\bf\tau}^1 +{\bf\tau}^2\otimes{\bf\tau}^2
+{\bf\tau}^1\otimes{\bf\tau}^2 )-
\]
\begin{equation}
-\frac{1}{2}({\bf\tau}^3\otimes{\bf\tau}^7 +{\bf\tau}^4\otimes{\bf\tau}^6
+{\bf\tau}^5\otimes{\bf\tau}^8 ),
\end{equation}
or, alternatively, (see (5.7))
\begin{equation}
{\cal G}=\frac{1}{4}Tr\left({\bf A}\otimes {\bf A}\right),
\end{equation}
where the 3$\times$3 matrix one--form ${\bf A}$ is defined as
\begin{equation}
{\bf A}=A_Bd \varphi^B = e_\mu {\bf\tau}^\mu
\end{equation}

In view of (5.11) the one--form $\bf A$ has a vanishing curvature
\begin{equation}
F_{BC}=A_{C,B}-A_{B,C}+ [A_B,A_C]=0,
\end{equation}
and thus $A_B$ is a pure gauge
\begin{equation}
A_B= (\partial_B U) U^{-1},
\end{equation}
where U is a 3$\times$3 matrix. To find it explicitly, it is convenient
to use Gauss decomposition of the general $SL(3,R)$ matrix
\begin{equation}
M=M_L M_D M_R,
\end{equation}
where $M_R, M_L$ are right-- (left--) triangle matrices and $M_D$ is
diagonal (all with unit determinant)
\begin{equation}
M_R= \left(\begin{array}{ccc}
1 & r_1 & r_3\\
0 & 1 & r_2\\
0 & 0 & 1\\
\end{array} \right),\;\;
M_D= \left(\begin{array}{ccc}
d_1 & 0 & 0\\
0 & d_2 & 0\\
0 & 0 & (d_1 d_2)^{-1}\\
\end{array} \right),\;\;
M_L= \left(\begin{array}{ccc}
1 & 0 & 0\\
l_1 & 1 & 0\\
l_3 & l_2 & 1\\
\end{array} \right).
\end{equation}

The corresponding generators are given by the sets $(e_3,\; e_4\; e_5)\quad,
(e_1,\; e_2)$ and $(e_6,\; e_7,\; e_8)$ respectively.

For any $U\in SL(3,R)/SO(3)$ one has $U_L=U_R^T,$ so using a parametrization
\begin{equation}
U_R= \left(\begin{array}{ccc}
1 & p_1 & p_3\\
0 & 1 & p_2\\
0 & 0 & 1\\
\end{array} \right),\quad
U_D= \left(\begin{array}{ccc}
q_1 & 0 & 0\\
0 & q_2 & 0\\
0 & 0 & (q_1 q_2)^{-1}\\
\end{array} \right),
\end{equation}
we obtain
\begin{equation}
U= \left(\begin{array}{ccc}
q_1 & q_1 p_1 & q_1 p_3\\
q_1 p_1 &q_2+q_1 p_1^2  &q_2 p_2+q_1 p_1 p_3\\
q_1 p_3 & q_2 p_2+q_1 p_1 p_3 &q_1 p_3^2+q_2 p_2^2+(q_1 q_2)^{-1}\\
\end{array} \right).
\end{equation}

Infinitesimal $SL(3,R)$ transformations
\begin{equation}
U\rightarrow M^T U M,\quad M=I+\lambda^{\mu} \hat{X}_{\mu}
\end{equation}
lead to the following representation of generators in terms
of the $q, p$--derivatives:
\[
\hat{X_1}=\frac{1}{3}(4 q_1 \partial_{q_1}-3 p_1 \partial_{p_1}
-2 q_2 \partial_{q_2}-3 p_3 \partial_{p_3}),
\]
\[
\hat{X_2}=\frac{1}{3}(2 q_1 \partial_{q_1}-3 p_2 \partial_{p_2}
+2 q_2 \partial_{q_2}-3 p_3 \partial_{p_3}),
\]
\[
\hat{X_3}=\partial_{p_1},
\]
\begin{equation}
\hat{X_4}=-p_1 \partial_{p_3} - \partial_{p_2},
\end{equation}
\[
\hat{X_5}=\partial_{p_3},
\]
\[
\hat{X_6}=2 q_2 p_2 \partial_{q_2}+ p_3 \partial_{p_1}+
\left(q_1^{-1}q_2^{-1}- p_2^2\right) \partial_{p_2},
\]
\[
\hat{X_7}=-2 q_1 p_1 \partial_{q_1}+ 2q_2 p_1 \partial_{q_2}-
\left(q_1^{-1}q_2- p_1^2\right) \partial_{p_1}+
\left(p_3- p_1p_2\right) \partial_{p_2}+
\left(p_1p_3- q_2 p_2 q_1^{-1}\right) \partial_{p_3}\;,
\]
\[
\hat{X_8}=-2 q_1 p_3 \partial_{q_1}+ 2q_2 p_1 p_2 \partial_{q_2}+
\left(p_1p_3- q_2 p_2 q_1^{-1}\right) \partial_{p_1}+ \]\[
\left(p_2p_3- p_1 p_2^2 + p_1 q_1^{-1}q_2^{-2}\right) \partial_{p_2}+
\left(p_3^2- q_2 p_2^2 q_1^{-1}- q_1^{-2}q_2^{-1}\right) \partial_{p_3} \;,
\]

Comparing this with our initial representation (4.1), (5.2), (5.3), (5.4)
in terms of the $\sigma$--model variables, one finds
\begin{equation}
q_1=-2 e^{2(\xi +2 \eta)/3},\;\;q_2= e^{2(\xi - \eta)/3},\;\;
p_1=u,\;\; p_2=-v,\;\; p_3=(\chi-uv)/2.
\end{equation}
As a result we obtain the following representation for the matrix
$U\in SL(3,R)/SO(3)$:
\[
U=-\mu
\left(
\begin{array}{ccc}
2 & 2a & \chi - av\\
2a & 2a^2-e^{-2\eta}& a(\chi - av) + ve^{-2\eta}\\
\chi-av & a(\chi-av)+ve^{-2\eta} &
\left((\chi -av)^2-2v^2e^{-2\eta}+ e^{-2(\xi + \eta)}\right)/2 \\
\end{array}
\right),
\]
\begin{equation}
\mu=\exp \left[ \frac{1}{3}(2\xi +4\eta )\right].
\end{equation}
(This matrix is related to one found by Maison \cite{ms} by some constant
$SL(3,R)$ transformation.)

Using (5.15) and (5.18) we obtain the metric of the target space
in terms of the coset $SL(3,R)/SO(3) $ variables, namely
\begin{equation}
{\cal G}=\frac{1}{4}Tr(dUU^{-1}\otimes dUU^{-1}).
\end{equation}
The equation of motion in these variables can be derived directly from the
action, or through the following argument \cite{egk}.
The pull--back of the Killing
one--form onto the configuration space
constitutes the set of Noether currents
\begin{equation}
J_i^\mu=\tau_A^\mu \;\frac{\partial\varphi^A}{\partial x^i}
\end{equation}
which are conserved in view of the invariance of the action under $SL(3,R)$:
\begin{equation}
\partial_i(h^{ij}\sqrt{h}J_i^\mu)=0.
\end{equation}
As it is known, for a symmetric target space such a set of conservation
laws is equivalent to the equations of motion.
Introducing the pull--back ${\cal A}={\cal A}_idx^i$ of {\bf A}, where
${\cal A}_i={ \rm A}_B \partial \varphi^B/\partial x^i$, we can rewrite (5.22)
as
\begin{equation}
d\star{\cal A}=0,
\end{equation}
or
\begin {equation}
 d\left(\star\; dUU^{-1}\right)=0,
\end{equation}
where the star stands for the 3-dimensional Hodge dual operation.
This form of the field equations is suitable for an aplication of the inverse
scattering transform method in the axisymmetric case \cite{bz}, \cite{egk}.
 An alternative development
following the prolongation structure technique can be found in \cite{mt}.
Another derivation of Belinskii--Zakharov type of a system
was given in \cite{br} using different
reduction scheme from the 5--dimensional KK theory.  Our purpose here was to
demonstrate explicitly how the $SL(3,R)$ symmetry, generally broken by the
non--critically coupled dilaton, turns out to be
restored for the critical coupling $\alpha=\sqrt{3}$.
Note, that another particular value of this coupling
constant, $\alpha =1$, which emerges in the context of the heterotic string
low--energy effective theory, belongs to the broken symmetry case.

\section{ Brans--Dicke--Maxwell model, $\alpha =0$ }
\renewcommand{\theequation}{6.\arabic{equation}}
\setcounter{equation}{0}

 For $\alpha =0$ the parametrization (3.3) of the target space
metric is not valid, and we have to restart with the initial parametrization
(2.25). Obviously, the dilaton decouples from the Einstei--Maxwell part of the
$\sigma$--model, which is described here in Neugebauer and Kramer variables
\cite{nk}.
As it is clear from (1.1), for $\alpha=0$  the system (including
the dilaton) is just the BDM
system with purely gravitationally coupled scalar field
(in the Einstein frame).
For completeness we present here the
corresponding Killing algebra in terms of real variables and then translate
it into the standard Ernst--potentials form.

Three of five  Killing vectors (3.35) from the maximal
solvable subalgebra of $sl(3,R)$, remain Killing vectors in the $\alpha=0$
case: $X_3,X_4,X_5$ (the electromagnetic and gravitational
gauge transformations).
The sum $ X_1+X_2$ is dilaton--independent and hence remains the symmetry
too (the scale transformation). The difference $X_1-X_2$ in the limit
$\alpha\rightarrow 0$ reduces to the pure dilaton shift.
In addition, a continuos duality rotation  emerges as a symmetry
(it is broken by the dilaton for any $\alpha \neq 0.$
The non--trivial part  of the target space isometry algebra consists of
the Harrison transformations
and the Ehlers transformation. Altogether one has the
9--dimensional algebra of
isometries generated by the following set of Killing vectors:
\begin{equation}
\begin{array}{l}
\bar X_1 =v\partial_a -a\partial_v,\\
\\
\bar X_2 =-(X_1+X_2)=2(f\partial_f +\chi\partial \chi)+
a\partial_a+v\partial_v \ ,  \\
\\
\bar X_3 =X_3, \quad \bar X_4 = X_4, \quad \bar X_5=X_5,\\
\\
\bar X_6=2fv\partial_f + \left(v\chi+aF-\right)\partial_\chi
+\left(\frac{1}{2}(v^2-3a^2)+f\right)
\partial_v + \chi_{-2}\partial_a ,\\
\\
\bar X_7 =2fa\partial_f +\left(a\chi - vF \right)\partial_f +
\left(\frac{1}{2}(a^2-3v^2))
+f\right)\partial_a - \chi_2\partial_v, \\
\\
\bar X_8 =2f\chi\partial_f + (\chi^2 -F^2)\partial_\chi + (v\chi -
aF)\partial_v + (a\chi + vF)\partial_a, \\
\\
\bar X_9=\partial_\phi ,
\end{array}
\end{equation}
where $F=f-(v^2+a^2)/2$.
 Obviously $\bar X_9$ commutes with all other generators, while the
remaining non--zero structure constants read
\[
{C^4}_{13}=-{C^3}_{14}=-{C^7}_{16}={C^6}_{17}=-{C^3}_{23}=-{C^4}_{24}={C^6}_{26}=
\]
\[
={C^7}_{27}=-{C^5}_{34}={C^2}_{37}={C^6}_{38}={C^2}_{46}=-{C^7}_{48}=1,
\]
\[
-{C^5}_{25}={C^8}_{28}={C^3}_{56}=-{C^4}_{57}={C^2}_{58}={C^8}_{67}=2,
\]
\begin{equation}
{C^1}_{36}=-{C^1}_{47}=3.
\end{equation}
The structure constants (6.2) form a $su(2,1)$ algebra. To cast it into  more
conventional form one has to introduce complex Ernst potentials
\begin{equation}
\varepsilon= f+ i\chi -\frac{v^2+a^2}{2},
\quad\Phi=\frac{v+ia}{\sqrt{2}}.
\end{equation}
For $\alpha =0$ the metric (2.25) in terms of these variables will read
\begin{equation}
{\cal G}=\frac{1}{2f^2}|d\varepsilon+2\Phi^*d\Phi |^2-\frac{2}{f}d\Phi
d\Phi^*+2d\phi^2.
\end{equation}

The complete isometry algebra of (6.4) is isomorphic to $su(2,1)\times R$. In
terms of the complex variables the generators (6.1) take the following form
\begin{equation}
\begin{array}{l}
\bar X_1= i\Phi\partial_\Phi + c.c. ,\\
\\
\bar X_2=2\varepsilon\partial_\varepsilon+ \Phi\partial_\Phi + c.c,\\
\\
\bar X_3=i(\frac{1}{\sqrt{2}}\partial_\Phi + \sqrt{2}
\Phi\partial_\varepsilon)+
c.c. , \\
\\
\bar X_4 = \frac{1}{\sqrt{2}}\partial_\Phi - \sqrt{2}\Phi\partial_\varepsilon +
c.c., \\
\\
\bar X_5= 2i\partial_\varepsilon + c.c. ,\\
\\
\bar X_6=\sqrt{2}\varepsilon\Phi\partial_\varepsilon+
\frac{1}{\sqrt{2}}(\varepsilon+2\Phi^2)\partial_\Phi + c.c. ,  \\
\\
\bar X_7= - i \sqrt{2}\varepsilon\Phi\partial_\varepsilon +
\frac{i}{\sqrt{2}}(\varepsilon -2\Phi^2)\partial_\Phi + c.c. , \\
\\
\bar X_8= -i\varepsilon(\Phi\partial_\Phi + \varepsilon\partial_\varepsilon)+
c.c. .
\end{array}
\end{equation}
Up to normalization, these generators coincide with ones given previously
by Neugebauer \cite{ne}, and Eris, G\"urses and Karasu \cite{egk}.
Contrary to the KK theory, which has natural description in terms of the
real target space variables, for the EM system more appropriate are
the complex variables which are
intrinsically related to nature of
the symmery group. It is interesting that inspite of this difference,
there is a striking
similarity between non--trivial sectors of the $sl(3,R)$ algebra
discussed in the previous section $(X_6,X_7,X_8)$ and $su(2,1)$ algebra here
$({\bar X}_6,{\bar X}_7,{\bar X}_8)$. One can easily see this
taking the corresponding commutators.

Five--dimensional target space for the BDM stationary system
is the product of a symmetric space $SU(2,1)/S(U(1)\times U(2))$ and
a line $R$. Hence the integrability arguments developed for EM system
equally apply to the present case. Nevertheless, both theories
are physically different. It is worth to be reminded once again
that the EM theory {\em is not} a particular case of
the EMD system since the dilaton equation (2.21) imposes
a constraint $F^2=0$ on the Maxwell field if only we put$\phi=const$.

\section{ Solution--generating technique for non--critical coupling}
\renewcommand{\theequation}{7.\arabic{equation}}
\setcounter{equation}{0}
{}From the above analysis it is clear that the stationary EMD system with
$\alpha\neq 0,\sqrt{3}$ does not possess enough symmetries to ensure
the existence of a full--scale solution generating technique similar to
that of the EM and KK theories. Nevertheless there are some non--trivial maps
between subspaces of the target space which can be used to generate
new solutions. These maps typically are descrete or present some
combinations of descrete maps and continuous transformations described
above. From five continuous transformations two are physically meaningful:
a constant dilaton shift, accompanied by suitable rescaling of
electromagnetic potentials,
\[
\phi_1 \rightarrow \phi_2= \phi_1+ \phi_{1c},
\]
\begin{equation} 
v_1 \rightarrow v_2=e^{\alpha \phi_{1c}}v_1,
\end{equation}
\[
a_1 \rightarrow a_2=e^{-\alpha \phi_{1c}}a_1,
\]
where $\phi_{1c}= {\rm const.}$, and the scale transformation
\[
f_1 \rightarrow f_2=C^2 f_1,\quad \chi_1 \rightarrow \chi_2=C^2\chi_1,
\]
\begin{equation}
v_1 \rightarrow v_2= C v_1,\quad  a_1 \rightarrow a_2= C a_1
\end{equation}
with the real constant $C$.

There are also some discrete symmetries of the target space valid for
all $\alpha$. One is discrete electric--magnetic
duality
\begin{equation}
\phi \rightarrow -\phi,\quad \chi \rightarrow -\chi,\quad v \leftrightarrow a,
\end{equation}
which obviously  leaves the metric (2.25) invariant.

Another discrete transformation is highly non--trivial. It is analogous
to the well--known
Bonnor transformation for the EM system \cite{bon}. To derive it,
let us consider two subspaces of the target space: the dilaton vacuum
\begin{equation}
dl_1^2=\frac{df^2+d\chi^2}{2f^2}+2d\phi^2,
\end{equation}
and the static magnetic sector
\begin{equation}
dl_2^2=\frac{df^2}{2f^2}-\frac{e^{2\alpha \phi}da^2}{f}+2d\phi^2.
\end{equation}
Under a (complex) transformation
\[
f_1^2=f_2e^{-2\alpha \phi_2},
\]
\begin{equation}
\chi_1=i\left(\frac{1+\alpha^2}{2}\right)^{1/2} a_2,
\end{equation}
\[
\phi_1=\frac{1}{2}\left(\phi_2+\frac{\alpha}{2}f_2\right),
\]
the line element (7.4) maps onto (7.5) up to a constant rescaling
\begin{equation}
dl_1^2=\left(\frac{1+\alpha^2}{4}\right) dl_2^2.
\end{equation}
Similarly, an ``electric'' transformation
\[
f_1^2=f_2e^{2\alpha \phi_2},
\]
\begin{equation}
\chi_1=i\left(\frac{1+\alpha^2}{2}\right)^{1/2} v_2,
\end{equation}
\[
\phi_1=\frac{1}{2}\left(\phi_2-\frac{\alpha}{2}f_2\right)
\]
maps (7.4) onto the line element of the electrostatic subspace
\begin{equation}
dl_2^2=\frac{df^2}{2f^2}-\frac{e^{-2\alpha \phi}dv^2}{f}+2d\phi^2,
\end{equation}
according the same rescaling rule (7.7). It is worth noting that, contrary
to the isometry transformations discussed in previous sections, as well
as to original Bonnor map, here we
deal with transfromations which  conformally rescale the target space
metric on  a constant factor. This does not change $\sigma$--model equations,
but does change the {\em three--metric}. If we restrict ourselves by the
axially--symmetric case, and represent the three--metric in
the Lewis--Papapetrou form
\begin{equation}
dl^2=e^{2\gamma} (d\rho^2+dz^2)+\rho^2d\varphi^2,
\end{equation}
from the three--dimensional Einstein equations we get
\begin{equation}
\gamma_2=\left(\frac{4}{1+\alpha^2}\right)\gamma_1.
\end{equation}

To apply these transformations for generating purposes one needs to know
appropriately complexified
seed solutions in the same way as in the original Bonnor case. To ensure
desired physical properties of resulting spacetime (e.g. asymptotic
flatness) one can use the dilaton shift (7.1)
and the scale transformation (7.2).
As an example we apply combination of (7.6) and (7.1) to the complexified
Kerr solution, which belongs to the subspace (7.4) with $\phi\equiv 0$,
\begin{equation}
ds^2=\frac{\Delta+b^2\sin^2\theta}{\Sigma}\left(dt-\omega d\varphi\right)^2-
\Sigma\left(\frac{dr^2}{\Delta}+d\theta^2+
\frac{\Delta\sin^2\theta}{\Delta+b^2\sin^2\theta}d\varphi^2\right).
\end{equation}
Here
\begin{equation}
\Delta=r^2-2mr-b^2,\quad \Sigma =r^2-b^2\cos^2\theta,\quad
\omega=-\frac{2imbr\sin^2\theta}{\Delta+b^2\sin^2\theta},
\end{equation}
and parameter $b$ is related to the original Kerr rotation parameter
$a$ as $b=ia$. The corresponding twist potential is pure
imaginary
\begin{equation}
\chi=2imb\cos\theta\Sigma^{-1},
\end{equation}
and the metric function $\gamma$ can be found from
\begin{equation}
e^{2\gamma}=\frac{P}{Q},\;\;
Q=\Delta+(m^2+b^2)\sin^2\theta,\;\;
P=\Delta+b^2\sin^2\theta.
\end{equation}

Let us apply to this solution a magnetic Bonnor--type transformation (7.6).
The resulting metric will read
\begin{equation}
ds^2=\left( \frac{P}{\Sigma}\right)^\nu dt^2-
\left( \frac{P\Sigma}{Q^2}\right)^\nu \frac{Q}{\Delta}(dr^2+\Delta d\theta^2)-
\left( \frac{\Sigma}{P}\right)^{\nu} \Delta \sin^2\theta d\varphi^2,
\end{equation}
where
\begin{equation}
\nu=\frac{2}{1+\alpha^2}.
\end{equation}
The corresponding magnetic potential and the dilaton function will  be
\begin{equation}
a=\frac{2\sqrt{\nu}mb\cos\theta}{\Sigma},\quad
\phi=-\frac{\alpha\nu}{2}\ln\frac{P}{\Sigma}.
\end{equation}
For $\alpha=0$ new solution reduces to the original Bonnor solution
for electrovacuum \cite{bon}.

Similarly, an electric transformation (7.8) applied to (7.12)
gives the same metric (7.16) and
\begin{equation} 
v=\frac{2\sqrt{\nu}mb\cos\theta}{\Sigma},\quad
\phi=\frac{\alpha\nu}{2}\ln\frac{P}{\Sigma},
\end{equation}
both solutions are related by the discrete duality (7.3).
A magnetic (electric) field coincides with that of a magnetic (electric)
dipole
\begin{equation}
p=\frac{2mb}{(1+\alpha^2)^{1/2}}.
\end{equation}
The Schwarzschild mass is given by
\begin{equation} 
M=\frac{2m}{1+\alpha^2},
\end{equation}
and the dilaton charge is related to mass as
\begin{equation}
D=\pm \alpha M.
\end{equation}
This relation is similar to that of the extremal
dilaton black holes \cite{ghs}.

It is worth noting that for KK case $\alpha=\sqrt{3}$ the target space
metric is not rescaled, as it is clear from Eq.~(7.7).
In this case the function $Q$ does not
enter the transformed solution (7.15). In the stringy case $\alpha=1$
the metric (7.16) is given
entirely in terms of rational functions.

\section{ Conclusion}
\renewcommand{\theequation}{8.\arabic{equation}}
\setcounter{equation}{0}

 By direct integration  of Killing equations for the
target space corresponding
to the stationary Einstein--Maxwell--dilaton theory we have shown
that for general  dilaton
coupling constant $\alpha$, with  notable exceptions of  $\alpha=0$ and
$\alpha =\sqrt{3}$, only  a five--dimensional solvable Killing algebra holds.
This algebra is isomorphic to
the maximal solvable subalgebra of the $sl(3,R)$ , to which  the
symmetryalgebra is enlarged in the exceptional case $\alpha =\sqrt{3}$. For
$\alpha\neq\sqrt{3}, 0$ the isometry group does not contain the essentially
non--trivial Ehlers--Harrison--type transformations and  does not
possess enough symmetries to render the system to be sharing
two--dimensional integrability property of the KK and BMD theories.

In two critical cases $\alpha=\sqrt{3}$ and $\alpha =0$ the target
space has the structure of cosets $SL(3,R)/SO(3)$ and
$(SU(2,1)/S(U(2)\times U(1))\times R$ respectively.
Both can be parametrized by five real variables in which the corresponding
lagrangians have
very similar structure. Using this formulation we have found
the non--trivial discrete (Bonnor--type)
transformations for an arbitrary dilaton coupling constant. Their
application (in combination with continuous transformations)
leads to new  solutions of dilaton gravity.

It may seem rather dissapointing that in the stringy case $\alpha =1$
the system does not possess enough symmetries to make the theory to be
two--dimensionally integrable.
This symmetry breaking by a non--critical dilaton may be attributed
to absence of the continuous duality rotation symmetry. This
is a special property of the truncated stringy gravity action (1.1)
without an axion field.  In fact the dilaton has to be considered rather as
representing the $SL(2,R)/U(1)$ coset \cite{mhs}.
Within this larger model the
continuous duality rotation {\em is} a symmetry, and consequently, the
isometry group of the corresponding target space is substantially enlarged
\cite{gk}.
\vskip3mm
{\bf Acknowledgement}\\

This work was supported in part by CONACYT, M\'exico. D.G. and
O.K. wish to thank the
Physics Department of the Centro de Investigacion y de Estudios Avanzados del
Instituto Polit\'ecnico Nacional for hospitality.
The work of D.G. and O.K.
was also supported in part by the ISF Grant M79000 and the Russian Foundation
for Fundamental Research Grant 93--02--16977.

\end{document}